\documentclass{aa}  

\usepackage{graphicx}
\usepackage{txfonts}
\usepackage{graphicx}
\usepackage{soul}

\usepackage[colorlinks=true, citecolor=blue]{hyperref}
\usepackage{subcaption}
\begin{document} 

\title{The bulge globular cluster Terzan 6 as seen from multi-conjugate adaptive optics and HST\thanks{Based on observations collected at the GEMINI South Observatory under program GS-2013A-Q23 (PI: Geisler), with the HST (GO14074, GO15616 and GO16420, PI:Cohen and Homan) and  at the Very Large Telescope of the European Southern Observatory at Cerro Paranal (Chile) under Program 091.D-0115 (PI:Ferraro)}}
   \author{M. Loriga
          \inst{1,2}
          \and
          C. Pallanca
          \inst{1,2}
          \and
          F.R. Ferraro 
          \inst{1,2}
          \and
          E. Dalessandro 
          \inst{2}
          \and
          B. Lanzoni 
          \inst{1,2}
          \and
          M. Cadelano 
          \inst{1,2}
          \and
          L. Origlia
          \inst{2}
          \and
          C. Fanelli
          \inst{2}
          \and
          D. Geisler
          \inst{3,4}
          \and S. Villanova
          \inst{5}
          }
   \institute{Dipartimento di Fisica e Astronomia, Università degli Studi di Bologna,
              Via Piero Gobetti 93/2, I-40129 Bologna
         \and
             INAF - Astrophysics and Space Science Observatory Bologna , 
             Via Piero Gobetti 93/3, I-40129 Bologna
             \and
     Departamento de Astronomia, Casilla 160-C, Universidad de Concepcion, Chile
     \and
Departamento de Astronomía, Facultad de Ciencias, Universidad de La Serena. Av. Juan Cisternas 1200, La Serena, Chile 
     \and 
     Universidad Andres Bello, Facultad de Ciencias Exactas, Departamento de Ciencias Físicas - Instituto de Astrofísica, Autopista
 Concepcion-Talcahuano 7100, Talcahuano, Chile
}

  \abstract
  {
  This work consists of the first detailed photometric study of Terzan 6, one of the least known globular clusters in the Galactic bulge. Through the analysis of high angular resolution and multi-wavelength data obtained from adaptive optics corrected and space observations, we built deep, optical and near-infrared color-magnitude diagrams reaching $\approx 4$ magnitudes below the main-sequence turnoff. Taking advantage of 4 different epochs of observations, we measured precise relative proper motions for a large sample of stars, from which cluster members have been solidly distinguished from Galactic field interlopers. A non-canonical reddening law (with $R_V=2.85$) and high-resolution differential reddening map, with color excess variations up to $\delta E(B-V) \approx 0.8 $ mag, have been derived in the direction of the system. According to these findings, new values of the extinction and distance modulus have been obtained: respectively, $E(B-V)=2.36\pm0.05$ and $(m-M)_0=14.46 \pm 0.10$ (corresponding to $d=7.8 \pm 0.3$ kpc). We also provide the first determinations of the cluster center and projected density profile from resolved star counts. The center is offset by more than $7\arcsec$ to the east from the literature value, and the structural parameters obtained from the King model fitting to the density profile indicate that Terzan 6 is in an advanced stage of its dynamical evolution, with a large value of the concentration parameter ($c = 1.94_{-0.26}^{+0.24}$) and a small core radius ($r_c = 2.6_{-0.7}^{+0.9}$ arcsec). We also determined the absolute age of the system, finding $t=13\pm 1 $ Gyr, in agreement with the old ages found for the globular clusters in the Galactic bulge. From the re-determination of the absolute magnitude of the red giant branch bump and the recent estimate of the cluster global metallicity, we find that Terzan 6 nicely matches the tight relation between these two parameters drawn by the Galactic globular cluster population.}
   
   \maketitle
%
\section{Introduction}
 
Galactic globular clusters (GCs) are populous and old stellar aggregates with typical ages of $\sim$12 Gyr \citep[see e.g][]{marin+09, vandenberg+13, valcin+20}. They are considered powerful tracers of the early phase of the Milky Way (MW) formation, and their study provides crucial insights into the evolutionary history of our galaxy.  Indeed, the recent exploration of the Galactic halo performed by the Gaia mission \citep{gaia+16} has demonstrated that the dominant fraction of GCs in the MW Halo has an extra-Galactic origin,  tracing the most relevant merging events (Gaia-Enceladus, Helmi stream, Sequoia, etc.) suffered by the MW over time. In this respect, the most famous example of a possible remnant of a remote merging event found in the Galactic halo is $\omega$ Centauri.  Despite its original classification of GC, this stellar system has been found to host multi-iron sub-populations \citep{norris+96, origlia+03, ferraro+04, bellini+09, villanova+14, bellini+17}, and its properties suggest that it is the remnant of a nuclear star cluster of an accreted dwarf galaxy \citep{bekki+03}. 

However, the star clusters in the central portion of our galaxy, the bulge, still remain largely unexplored, due to the large extinction and crowding in its direction. However, the bulge contains about $25\%$ of the total stellar mass of the MW and represents the first massive structure to have formed. Hence, understanding its structure and evolution and constraining the properties of its stellar populations is key to describing the formation and early evolutionary processes of the in situ Galaxy.  
Thanks to the recent development of the adaptive optics (AO) techniques in the near-infrared (NIR), a huge improvement in the knowledge of bulge GCs has been obtained \citep[see e.g][]{saracino+15,saracino+19}. In this respect, recent studies \citep{ferraro+09, ferraro+16, ferraro+21}, have shown that two stellar systems traditionally cataloged as bulge GCs, namely Terzan 5 and Liller 1, host populations with remarkable differences in age ($\Delta t$ up to $\sim 10-11$ Gyr) and in iron abundance ($\Delta$[Fe/H]$\sim 1$ dex), and for these reasons, they might be the fossil remnants of more massive structures that contributed to the formation of the bulge (see also \citealp{lanzoni+10, origlia+11}, \citeyear{origlia+13}, \citeyear{origlia+19}; \citealp[]{massari+14},\citeyear{massari+15}; \citealp[]{pallanca+21, dalessandro+22, crociati+23, deimer+24, fanelli+24a}). The discovery of their nature has further emphasized the urgency of an appropriate exploration of the GC population in the Galactic bulge. 

\begin{figure}
   \centering
   \includegraphics[width=9cm]{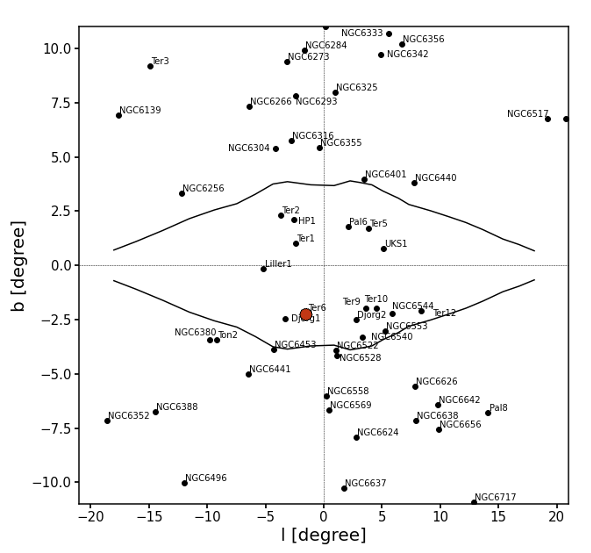}
      \caption{Position in Galactic longitude (l) and latitude (b) of the known GCs in the direction of the bulge (from the \citealt{harris+96} compilation). For reference, the black lines represent the outline of the inner bulge. The position of Terzan 6 is shown as a large red circle.}
    \label{GCs_map}
\end{figure}

In this exciting context, here we present a multi-wavelength investigation of the poorly studied bulge GC Terzan 6, discovered by \citet{terzan+68} using Schmidt plates obtained at the Haute-Provence Observatory.  It is located at longitude l$=358.571^{\circ}$ and latitude b$=-2.162^{\circ}$ \citep{barbuy+97}, at a distance of 1.3 kpc from the Galactic center and approximately 7 kpc from Earth \citep[][see also \citealp{fahlman+95, barbuy+97, baumgardt+21}]{valenti+07}. As illustrated in Fig. \ref{GCs_map}, it is located in the inner bulge, where the mean color excess is as high as $E(B-V)=2.35$ \citep{harris+96}, and this is the main reason why it has been poorly investigated so far.
Terzan 6 has an intermediate integrated luminosity ($V$-band absolute magnitude $M_V = -7.56$; \citealp{harris+96}), with an estimated total mass of about $10^5 M_\odot$ \citep{baumgardt+18}, and it shows a very concentrated structure, suggesting that it is a possible post-core collapse system \citep{trager+95}. It hosts two low-mass X-ray binaries that exhibit Type-I X-ray bursts and sharp eclipse signals during outburst phases \citep[see][and references therein]{painter+24, vandenberg+24}. While NIR low-resolution spectroscopy of CaII triplet (CaT) lines suggests [Fe/H]$=-0.21\pm 0.15$ \citep{geisler+23}, recent high-resolution spectroscopic analysis of a sample of giants \citep{fanelli+24b} shows a metallicity of [Fe/H]$=-0.65$ for the stellar population hosted in the cluster. 

Here we present the most detailed study so far of the stellar population hosted in Terzan 6, taking advantage of a combination of photometric data acquired through adaptive optics-assisted, ground-based facilities, and space-based instruments, which allowed the acquisition of images of superb quality and spatial resolution. 
This work is organized as follows. Section \ref{Observations and Data Analysis} describes the optical and the NIR data sets used in the analysis, and the adopted data analysis procedures. In Section \ref{NIR and Optical CMDs of Terzan6} we present the derived optical, NIR, and hybrid color-magnitude diagrams (CMDs), and discuss the main characteristics of evolutionary sequences. Section \ref{Proper motion analysis} presents the relative proper motion (PM) analysis and the decontamination procedure adopted to remove the Galactic field interlopers, while Section \ref{Differential reddening} describes the method adopted to evaluate the extinction law and the differential reddening map in the direction of the cluster. In section \ref{Determination of the structural parameters} we present a new determination of the center of gravity, projected density profile, and main structural parameters of the system, obtained from resolved star counts and King model fitting.  Section \ref{Distance and Age determination} is focused on the determination of the distance modulus of Terzan 6 through the comparison with the CMD of the well-studied GC NGC 6624, and the estimate of its absolute age through isochrone fitting. We also show the position of Terzan 6 in the age-metallicity distribution drawn by bulge GCs. Section \ref{sec:bump} is devoted to the determination of the absolute magnitude of the red giant branch (RGB) bump of the system. Finally in Section \ref{Conclusions} we present a summary of the work and the main conclusions.

\section{Observations and data analysis}\label{Observations and Data Analysis}
   \subsection{Optical and NIR data sets}\label{Optical and NIR Data Sets}
    \begin{table*}
   \caption{Summary of the data sets used}             
   \label{table:1}      
   \centering          
   \begin{tabular}{c c c c l c}    
   \hline\hline 
    Instrument & Program ID & PI & Date & Filter & $N_{exp}$ x $t_{exp}$\\
    &   &   &  (yyyy/mm/dd) & & \\
   \hline   
  \rule{0pt}{0.5pt} \\
   HST/ACS-WFC & 14074 & Cohen & 2016 July 24 & F606W & 4 x 494s \\  \\
   HST/ACS-WFC & 15616 & Homan & 2019 June 29 & F606W & 4 x 361s \\
   & & & & F814W & 4 x 375s \\ \\
   HST/ACS-WFC & 16420 & Homan & 2021 November 11 & F606W & 4 x 480s\\
   & & & & F814W & 4 x 473s \\ \\
   GEMINI/GeMS-GSAOI & GS-2013A-Q23 & Geisler & 2013 June 8-12 & J & 23 x 30s \\
   & & & & $K_s$ & 20 x 30s \\ \\
   ESO/HAWK-I & 091.D-0115(B) & Ferraro & 2013 April 21 & J & 2 x 320s \\
   & & & & $K_s$ & 1 x 240s \\
   \hline                  
   \end{tabular}
   \end{table*}

\begin{figure*}
    \centering
    \includegraphics[width=18cm]{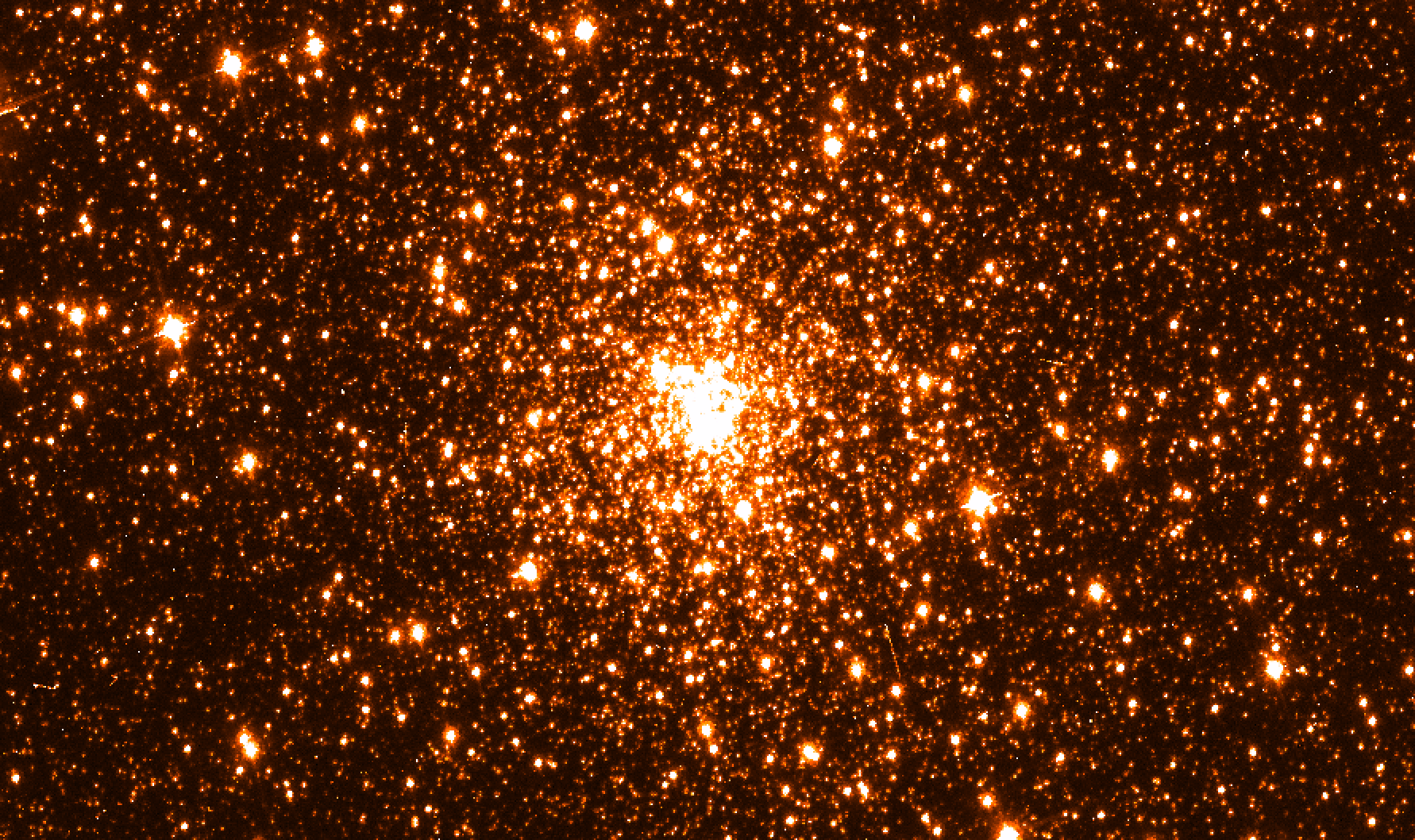}
    \caption{A ($60\arcsec \times 35\arcsec$) portion of a WFC/ACS image centered on Terzan 6 obtained in F814W with an exposure time of 473s. North is up, East is to the left.}
    \label{Ter6_wfc}%
\end{figure*}

The photometric investigation of Terzan 6 presented in this study combines optical and NIR datasets. In the following we schematically present the adopted datasets.   
The optical photometric data set is composed of high-resolution, multi-epoch images obtained from the Wide Field Channel (WFC) of the Advanced Camera for Survey (ACS) on board Hubble Space Telescope (HST). The WFC/ACS provides $\sim$ 0.05 arcsec/pixel spatial resolution in a nominal $202 \times 202$ arcsec$^2$ field of view (FoV).
In particular, the analyzed optical data set is composed of images acquired in three different epochs. 
The first epoch was acquired in 2016 (Proposal ID. 14074 P.I Cohen) and comprises images obtained only in the F606W filter. The second one was performed in 2019 (Proposal ID. 15616 P.I. J. Homan), while the third one in 2021 (Proposal ID. 16420 P.I. J. Homan). They are both composed of images obtained in the F606W and F814W filters. For the sake of illustration, Fig.~\ref{Ter6_wfc} shows the portion of an image (in the F814W filter) containing the central region of the cluster.
    
\begin{figure}
    \centering
    \includegraphics[width=7cm]{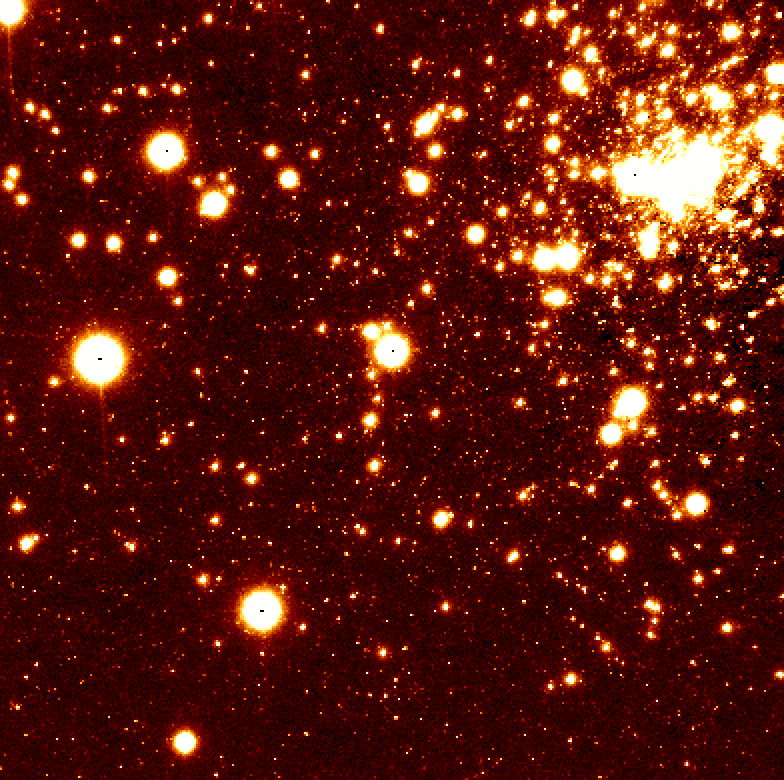}
    \caption{A ($40\arcsec \times 40\arcsec$) image of Terzan 6 obtained in $K_s$ filter of GSAOI with an exposure time of 30s. North is up, East is to the left.}
    \label{gemini_chip4}%
\end{figure}

The NIR data set consists of a set of high-resolution images obtained in 2013 with the Gemini South Adaptive Optics Imager (GSAOI) assisted by the Gemini Multi-Conjugate Adaptive Optics System (GeMS), mounted at the 8m Gemini South Telescope.
The GSAOI camera is a 4k x 4k NIR imager covering 85$\arcsec \times 85\arcsec$ designed to work at the diffraction limit of the 8-meter telescope with a spatial sampling of 0.02$\arcsec$/pixel. The analyzed data set is part of the GEMINI program GS-2013A-Q23 (P.I D. Geisler) and it is composed of a set of 23 images in the J filter and 20 in the $K_s$ filter.
Figure \ref{gemini_chip4} shows a $K_s$-band image of the GSAOI chip containing the central region of cluster.

For photometric calibration purposes, the Gemini images were complemented by HAWK-I (High Acuity Wide field K-band Imager) observations in the K and J bands acquired in 2013 (Proposal ID. 091.D-0115(B) P.I F.R. Ferraro) at the Very Large Telescope. HAWK-I  is a wide-field NIR imager that covers $7.5\arcmin \times 7.5\arcmin$ in the sky with a pixel scale of 0.106$\arcsec$/pixel. 

\begin{figure}
   \centering
   \includegraphics[width=9cm]{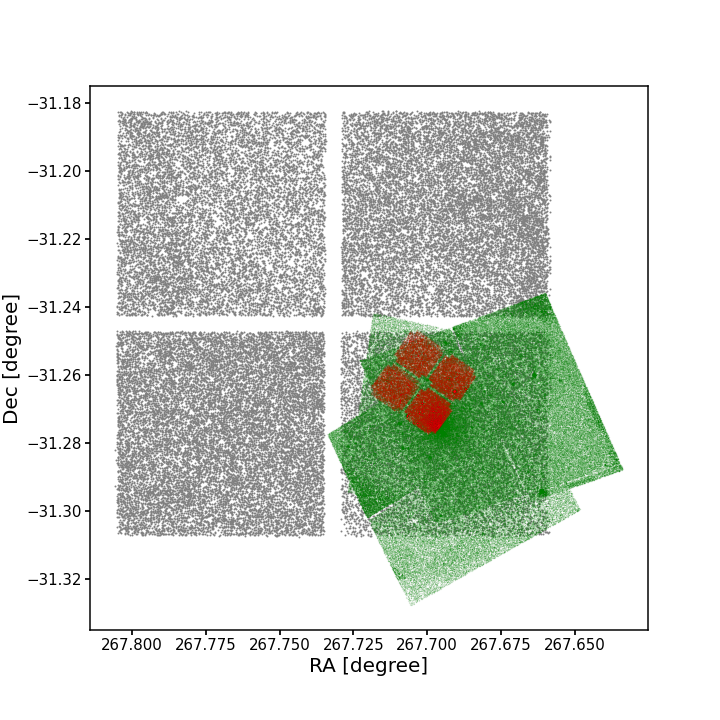}
   \caption{FoVs of the data sets analyzed in this study: the HAWK-I's FoV ($7.5\arcmin \times 7.5\arcmin$) is plotted in gray, the FoV of the different HST/WFC/ACS pointings in green, and that of GeMS is in red.}
   \label{mappa_hawki}
   \end{figure}

Fig.~\ref{mappa_hawki} shows the FoV covered by the different datasets and Table~\ref{table:1} summarizes all the exposures analyzed in the present study.

\subsection{Data reduction}\label{Data Reduction}   
For the NIR data set, we performed the pre-reduction procedure individually for each of the four chips, by using standard \texttt{IRAF}\footnote{The Image Reduction and Analysis Facility is distributed by the National Optical Astronomy Observatories (NOAO), which is operated by the Association of Universities for Research in Astronomy, Inc., under a cooperative agreement with the National Science Foundation.} tools to correct for bias and flat field and to perform the sky subtraction (\texttt{imcombine} and \texttt{flatcombine} to combine the different sky and flat images respectively, \texttt{ccdproc} to apply the correction to the different scientific images). To estimate the unresolved background we combined the target scientific images in order to build a MASTER SKY. For the images secured with HAWK-I the pre-reduction procedure has been performed by using the \texttt{EsoReflex} software \citep{freudling+13}, with the specific HAWK-I pipeline. The evaluated full-width at half maximum (FWHM) is $\approx 400$ mas for the $K$-band image and $\approx 450$ mas for the J ones.

As for the HST data, we used $\_$flc images, which are already calibrated and corrected for the Charge Transfer Efficiency (CTE). In addition these images are already processed by the Space Telescope Institute dedicated pipeline and are corrected for bias and flat-fielding effects. We applied the Pixel Area Map corrections by exploiting the time-averaged files available on the ACS website\footnote{\href{https://www.stsci.edu/hst/instrumentation/acs}{www.stsci.edu/hst/instrumentation/acs}}.

For both optical and NIR data the photometric reduction procedure was performed independently for each chip, filter and epoch. The photometric analysis of the images was performed by point spread function (PSF)-fitting since this approach allows the derivation of accurate measures of the source luminosities even in the case of a high level of crowding, as in the case of the central regions of a GC. To this aim, we made use of the \texttt{DAOPHOT} \citep{stetson+87} software. We used a sample of bright, isolated and uniformly distributed stars to model the PSF using the \texttt{DAOPHOT/PSF} routine and allowing a cubic polynomial spatial variation. The analysis of the PSF for the NIR dataset allowed us to perform a quality check of all the acquired images. Thus, the 18 J-band images (with mean FWHM $\approx 150 $ mas) and 13 $K_s$-band images (with mean FWHM $\approx 80$ mas) acquired in the best observational conditions have been selected for the PSF fitting analysis.
The obtained PSF models were then applied to all star-like sources detected at $4 \sigma$ (for GEMINI and HAWK-I) and $5 \sigma$ (for HST) above the background level using \texttt{ALLSTAR}.
At this point, for each image, we had a list with instrumental coordinates, magnitudes and relative errors for the detected stars. We used it to create a combined master star list for HST and GEMINI, which included all the stars measured in at least one image. Following the approach of previous studies \citep[e.g][and references therein]{dalessandro+14}, this master list was used as input for \texttt{ALLFRAME} \citep{stetson+94}, which was performed independently for each chip and epoch. The resulting output files were then combined to generate a catalog containing all the stars detected in more than half of the images (per chip and filter), using as reference the WFC1 HST's 2021 epoch, which covers the cluster's center. This criterion removed the vast majority of spurious detections (such as those due to cosmic rays etc). For each stellar source in the catalogs, the various magnitude estimates were homogenized. The mean values and standard deviations of these estimates were then assigned as the star magnitudes and photometric errors in the final catalog (\citealp{ferraro+91},\citeyear{ferraro+92}).
   
\subsection{Astrometry and calibration}\label{Astrometry and Calibration}
The next necessary step to build the final catalog was to transform the instrumental coordinates to the absolute reference frame ($\alpha$ and $\delta$) and to perform the photometric calibration of the magnitudes. First of all, it was essential to correct the instrumental coordinates for the effects of geometric distortions. To this aim, we followed \citet{bellini+11} for ACS-WFC, while for the GeMS/GSAOI camera, we applied the solutions provided by \citet{dalessandro+16}. 
The astrometric transformation of the catalog has been performed by using the stars in common with the publicly available Gaia DR3 catalog \citep{gaia+23} and through the cross-correlation software \texttt{CataXcorr} \citep{montegriffo+95}.

HST magnitudes were reported to the VEGAMAG photometric system, utilizing the zero points and the encircled energy fraction provided on the ACS website for a 1$\arcsec$ (20 pixels) aperture and performing the aperture correction. The J and $K_{s}$ instrumental magnitudes have been reported to the 2MASS photometric system by using the Vista Variables in the Via Lactea (VVV) catalog \citep{smith+18}. However, since the number of stars in common between the GEMINI catalog and VVV is quite low, we used the ESO/HAWK-I images 
to derive a sort of  "bridge catalog" to report the J and $K_s$ magnitude of the GEMINI catalog in the VVV system.
As shown in Fig.~\ref{mappa_hawki} the GeMS FoV is entirely contained in just one HAWK-I chip (Chip 2), thus the HAWK-I catalog has been cross-correlated with the VVV one and finally used to calibrate the GeMS magnitudes. 
\subsection{NIR and Optical CMDs of Terzan 6}\label{NIR and Optical CMDs of Terzan6}

\begin{figure*}
   \centering
 \includegraphics[width=19cm]{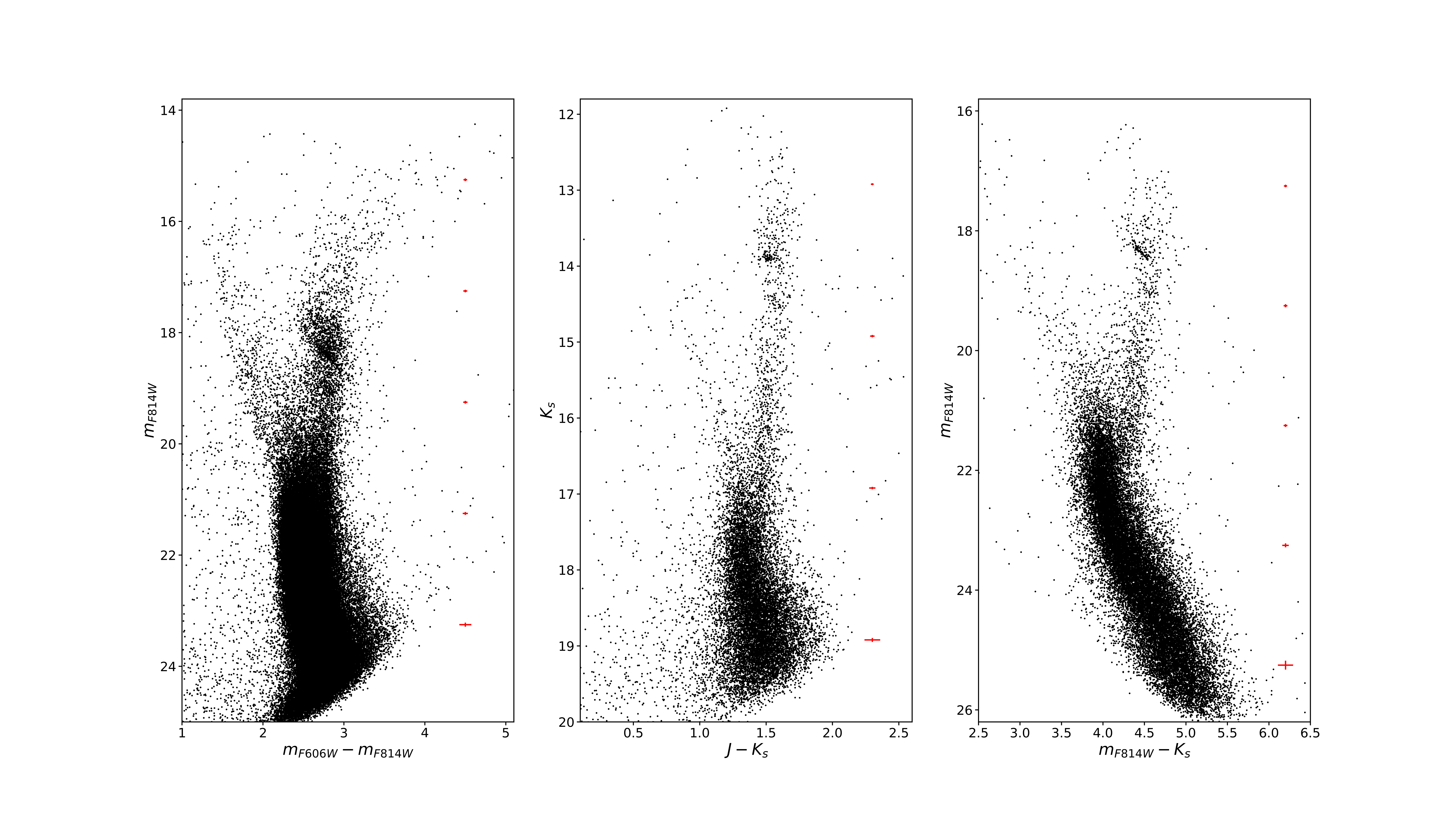}
   \caption{($m_{F814W}$, $m_{F606W}-m_{F814W}$), ($K_s$, $J-K_s$) and ($m_{F814W}$, $m_{F814W}-K_{s}$) observed CMDs of Terzan 6 obtained from the analysis of the HST and GeMS images discussed in the text. The small red crosses in the right-hand side of each panel show the photomeric errors at different magnitude levels.}
    \label{NIR_cmd}
    \end{figure*}

Fig.~\ref{NIR_cmd} shows the CMDs obtained from the analyzed observations in different filter combinations. In particular panel (a) shows the   ($m_{F814W}$, $m_{F606W}-m_{F814W}$) optical CMD, panel (b) the ($K_s$, J-$K_s$)-CMD in the pure NIR bands and panel (c) the hybrid optical-NIR ($m_{F814W}$, $m_{F814W}-K_s$) CMD.  The evolutionary sequences can be well identified in all the diagrams. It is possible to distinguish the red clump at $K_s \approx 14$ ($J \approx 15.5$, $m_{F814W} \approx 18.5$, $m_{F606W} \approx 21$) a well-extended RGB with the RGB-bump clearly visible at $K_s \approx 14.5$ and $m_{F814W} \approx 19$ and the Sub-Giant Branch (SGB). The CMD extends to more than $\approx$ 4 mag below the Main Sequence Turn Off (MS-TO), which is clearly visible at $m_{F814W} \approx 22$ ($K_s \approx 17.5$).

The comparison of the CMDs shown in Fig.~\ref{NIR_cmd} with previous results published in the literature (\citealp[Figure 2 of][]{fahlman+95}; \citealp[Figure 2 of][]{valenti+07}) allows to fully appreciate the advantages of the use of multi-conjugate adaptive optic systems (like the GeMS/GSAOI system), which are able to provide results comparable to those obtainable from space (see \citealp[Figure 18 of][]{cohen+18}). 

\section{Proper motion analysis}\label{Proper motion analysis}
\begin{figure*}
   \centering
  \includegraphics[width=15cm]{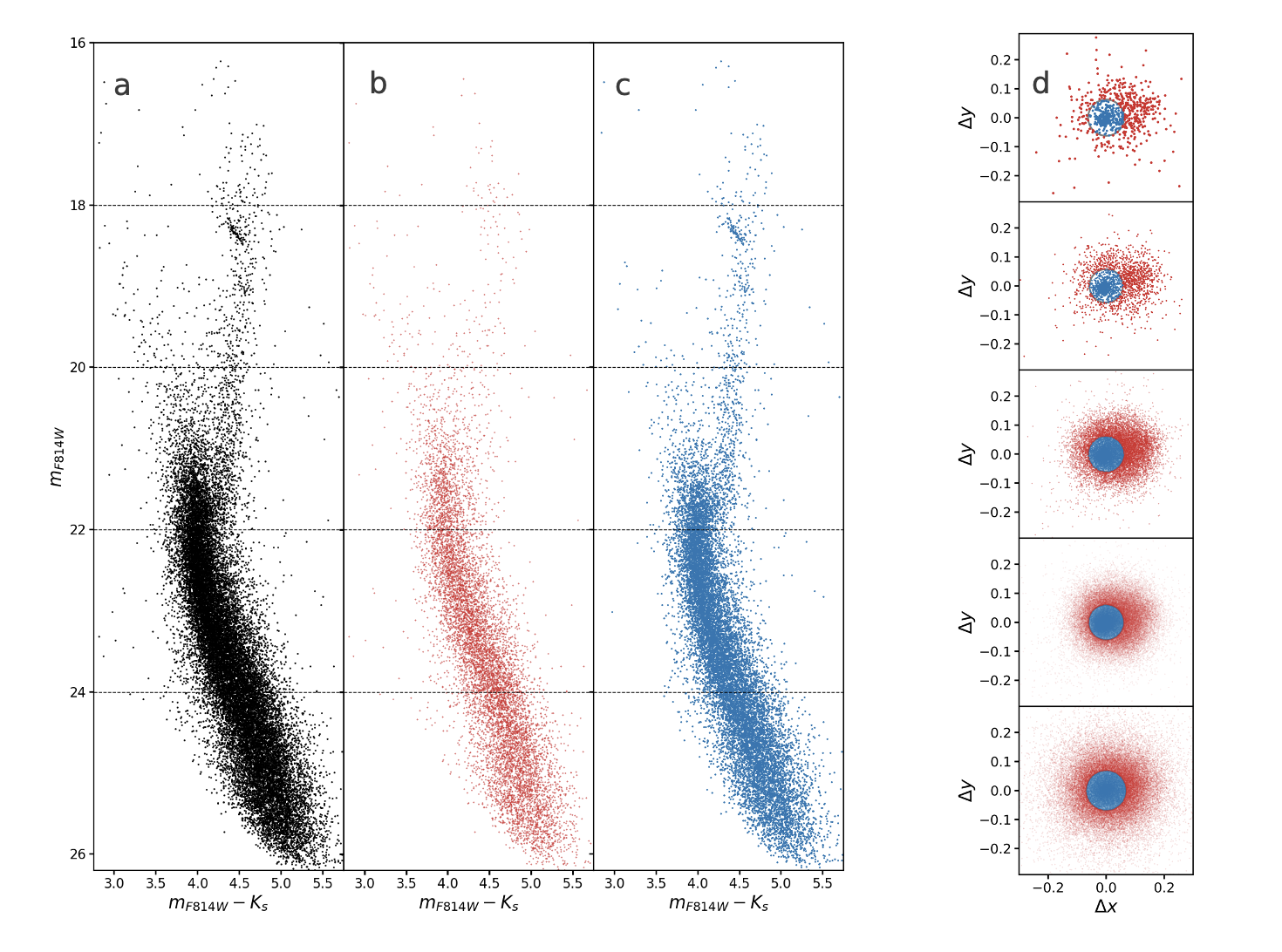}
   \caption{Panel a: Hybrid CMD of all the stars in common between GEMINI and at least one of the HST epochs. Panel b: CMD of the likely field interlopers according to the PM analysis. Panel c: CMD of the likely cluster members according to the measured PMs. Panel d: VPDs of the measured stars. The circles have a radius equal to $3\sigma$, with $\sigma$ being the average PM error in each magnitude bin. The blue (red) dots within (beyond) the circles are classified as cluster members (field interlopers) and their CMD is plotted in panel c (panel b).}
    \label{proper_motions_vpd}
\end{figure*}

The characterization of the evolutionary sequences requires an appropriate decontamination of the CMD from galactic field interlopers. This is particularly critical in the case of Bulge GCs because they are heavily embedded into the Bulge, thus a non-negligible fraction of the sources detected along the line-of-sight are likely field stars. In particular, we exploited the large temporal baseline provided by the multi-epoch data set (see Table~\ref{table:1}) to perform an accurate relative PM analysis. We used different temporal baselines depending on how many (2,3 or 4) observations of the same star are available in the datasets. For this analysis we used the approach described in \citet[see also \citealp{massari+13,bellini+14,massari+15,cadelano+22,cadelano+23}]{dalessandro+13} and to avoid saturation problems we considered stars with $m_{F814W}>16$. Briefly, the procedure consists of determining the displacement of the centroids of the stars measured in two or more epochs once a common coordinate reference frame is defined. The first step is to adopt a distortion-free reference frame, hereafter "master frame". The master-frame catalog contains stars measured in all the ACS/WFC F814W-band single exposures secured in 2021 (third HST epoch). To derive accurate geometrical transformation between our catalog and the master frame we selected a sample of $\approx 3000 $ bona-fide stars uniformly distributed in the FoV and distributed along the RGB and the red clump at magnitudes $17<m_{F814W}<20$ and colors $2.50<(m_{F606W}-m_{F814W})<3.50$, to maximize the membership probability.  
The mean position of a single star in each epoch has been measured as the ($3\sigma-$)clipped mean position calculated from all the N individual single-frame measurements. The relative rms of the position residuals around the mean value divided by N has been used as the associated error ($\sigma$).
Finally, the displacements are obtained as the positional differences among the different epochs. The error associated with the displacement is the combination of the errors on the positions in each epoch.
The relative PMs ($\mu$x, $\mu$y) are finally determined by measuring the difference of the mean x and y positions of the same stars in the different epochs, divided by their temporal baseline $\Delta$t. Such displacements are in units of pixels $yr^{-1}$. In the case of 3 or more epochs, we evaluated the PMs as the angular coefficients of the relation between the x and y positions and the time, expressed in Julian Day (JD). \\
We then iterated this procedure a few times by removing likely non-member stars from the master reference frame based on the preliminary PMs obtained in the previous iterations.
The convergence is assumed when the number of reference stars that undergo this selection changes by less than $ \approx 10\%$ between two subsequent steps.\\
As result, we derived relative PMs for 135134 stars.
In particular, we computed relative PMs by using only 2 different epochs for 71201 stars; while for 53118 stars we exploited 3 different epochs, and finally 10815 stars were observed in 4 different epochs. To build a clean sample of stars with a high membership probability, we constructed the vector-point-diagram (VPD) in 5 magnitude bins with a width of 2 mag each (see Fig.~\ref{proper_motions_vpd}d)
As expected, the PM distributions get broader for increasing magnitudes because of the increasing uncertainties in the centroid positions of faint stars. By definition in each magnitude bin, cluster stars are expected to be distributed around position (0,0), and we adopted as likely cluster members all the stars included within 3$\sigma$ from this value, where $\sigma$ is the average PM error in that magnitude bin. The CMD of the member stars selected according to this criterium is shown in Fig.~\ref{proper_motions_vpd}c, while that of Galactic field interlopers is plotted in Fig.~\ref{proper_motions_vpd}b. 
As clearly visible (especially in the top panels of Fig.~\ref{proper_motions_vpd}d), the PM  of the bulge field and disk stellar populations describe a vaguely elliptical broad distribution extending toward positive values of $\Delta x$ and $\Delta y$. This distribution partially overlaps the portion of the VPD where cluster members are selected, thus some residual contamination is expected in the PM-cleaned CMDs. Nevertheless, we can notice that all the evolutionary sequences are much better defined in the PM-cleaned CMD (Fig.~\ref{proper_motions_vpd}c). 
This also demonstrates that ground-based AO observations can be successfully combined with HST observations to perform a reliable PM analysis even in dense environments \citep[see also][]{saracino+19,ferraro+21,dalessandro+22}.

\section{Differential reddening}
\label{Differential reddening}

Because of the position of Terzan 6 in the Bulge of the Galaxy, at only about 1 kpc from the Galactic center, one of the most severe challenges in its characterization is represented by the huge and differential interstellar absorption along the line of sight, due to the presence of clouds with different column densities on scales that can be as small as a few arcseconds \citep[see, e.g.,][]{pallanca+21}. 
This implies that, depending on its location in the acquired image, each star is affected by a different color excess $E(B-V)$, which is defined as the difference between the observed color $(B-V)$ and the intrinsic one $(B-V)_0$. The main resulting effect on the CMD is an elongation of the main evolutionary features along the direction of the reddening vector, an effect that is commonly referred to as differential reddening.  To correct the observed sample for differential reddening we proceeded in two steps: STEP1 - the identification of the direction of the reddening vector in the adopted CMDs; STEP2 - the operative star-by-star correction.
 
STEP1 $-$ The direction of the reddening vector depends (through the wavelength-dependent parameters commonly named $R_\lambda$) on the photometric filters used to build the CMD. These parameters are linked to the extinction coefficient $A_{\lambda}$ through  $A_{\lambda}=R_{\lambda}\times E(B-V)$, and they can be expressed as the product of two terms: $R_\lambda = R_V \times c_{\lambda, R_V}$. The coefficient $c_{\lambda, R_V}$ is therefore equal to the ratios $A_{\lambda}/A_V$ and $R_{\lambda}/R_V$, which express the wavelength dependence of interstellar extinction relative to the absolute extinction in the $V$-band, and are commonly referred to as ``reddening law" \citep[see, e.g.,][]{cardelli+89,fitzpatrick+90,odonnell+94,fitzpatrick+99}. The $V$-band extinction coefficient $R_V$ is usually assumed to be 3.1, which is the standard value for diffuse interstellar medium \citep{schultz+75,sneden+78}, but it has been observed to vary along different directions toward the inner Galaxy \citep[][and references therein]{popowski+2000,nataf+13b,casagrande+14,alonso+17,pallanca+21}. This implies that the reddening law can be steeper or shallower along different lines of sight, because the wavelength dependence of $c_{\lambda, R_V}$ (or  $A_\lambda/A_V$, $R_\lambda/R_V$) changes  with the adopted value of $R_V$ (see, e.g., Fig.3 of \citealp{cardelli+89} or Fig.3 in \citealp{pallanca+21}). Thus, to determine the value of $R_V$ along the line of sight of Terzan 6, we selected a sample of stars along the bluest side of the RGB in the optical CMD, and another sample along the reddest RGB envelope. Then, we adopted different values of $R_V$ and, for each of them, we shifted the red sample along the corresponding direction of the reddening vector until it matched the blue sample. This has been done in both the optical and the NIR CMDs, each time measuring the amount of shift needed to superpose the two samples. We found that the best superposition of the two samples in the optical and the NIR CMDs \textit{simultaneously}, 
implying approximately the same values of $E(B-V)$ needed in the two shifts, 
is obtained\footnote{We remark that slightly different extinction laws cannot be solidly excluded by the available data.} for $R_V = 2.85$. The corresponding values of $R_{\lambda}$ in the photometric filters of interest for our study therefore are: $R_{\rm F606W}= 2.64$, $R_{\rm F814W}=1.70$, $R_{J}=0.79$, and $R_{K_s}=0.32$.

STEP2 $-$ Once the appropriate extinction law has been set, to correct the CMDs we adopted the so-called star-by-star approach, which is described in detail by \citet[see also \citealp{cadelano+20a}]{pallanca+19}. In summary, in the optical HST CMD we first determined the mean ridge line (MRL) of Terzan 6 using a sample of well-measured and likely member stars located along the RGB, SGB and bright MS. Then, for each star in our HST catalog, we selected a group of nearby (within a limiting radius of $10\arcsec$), well-measured stars ($N_{*}$) to create a ``local CMD". By progressively shifting the MRL along the direction of the reddening vector, in steps of $\delta E(B-V)$, we evaluated the residual color $\Delta VI$ as 
\begin{equation}
    \Delta VI = \sum^{N_*}_{i=1}(|VI_{obs,i}-VI_{MRL,i}| + w_i\times |VI_{obs,i}-VI_{MRL,i}|)
\end{equation}
where $VI_{obs,i}$ is the $(V-I)$ color for each of the $N_*$ stars of the local CMD, $VI_{MRL,i}$ is that of the MRL at the same level of magnitude, and the weight $w_i$ depends on both the photometric error on the color ($\sigma$) and the spatial distance ($d$) of the $i$th source from the considered star. So the final value of $\delta E(B-V)$ assigned to each star is the one which minimizes the residual color $(\Delta VI / N_*)$.

\begin{figure*}
   \centering
    \includegraphics[width=15cm]{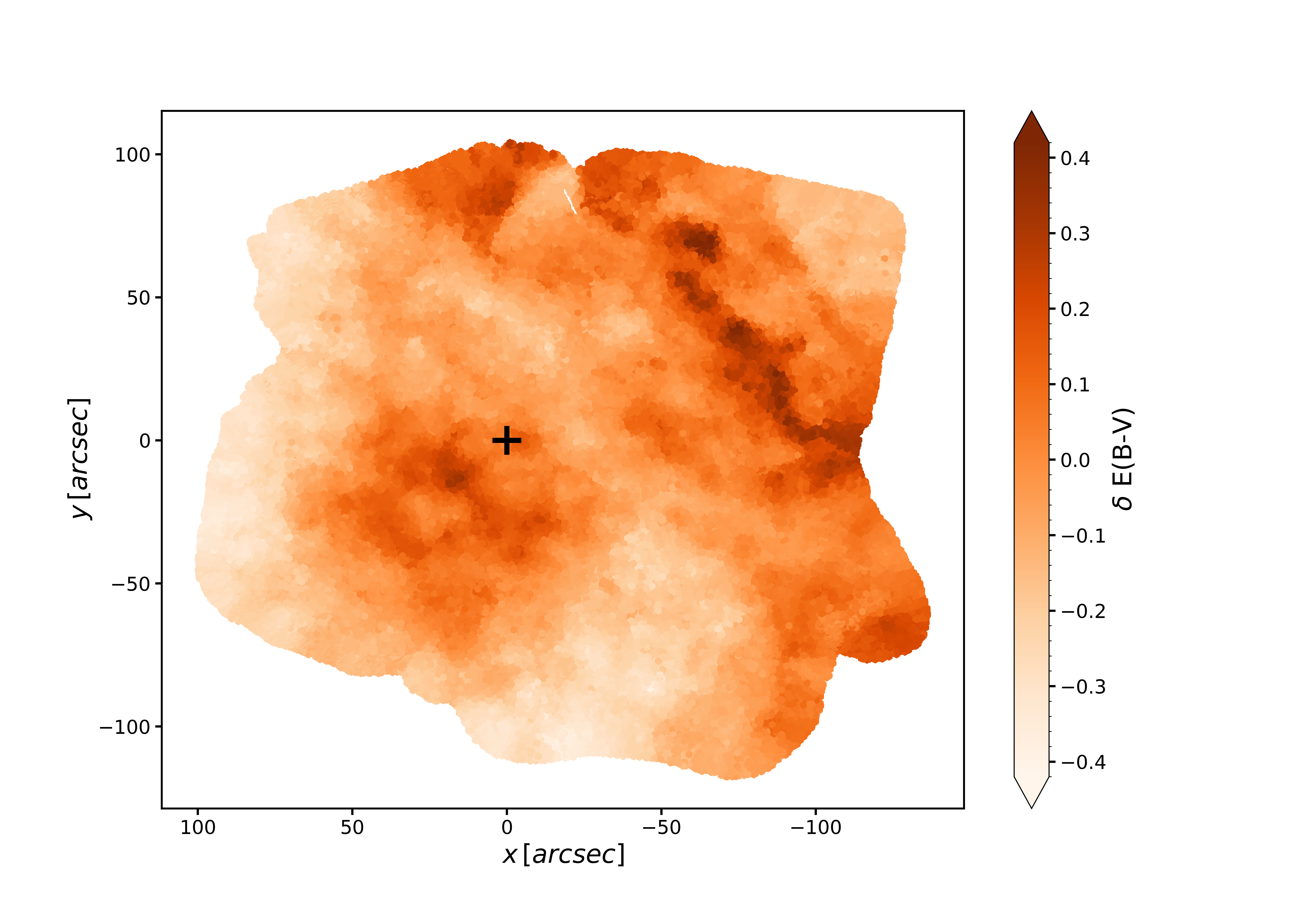}
   \caption{Differential reddening map relative to the cluster center position (black cross, see Sec.\ref{Determination of the structural parameters}) of the portion of the HST FoV where it was possible to derive a reliable correction. North is up, East is to the left. Darker colors correspond to more extincted regions, as detailed in the side color bar.}
    \label{reddening_map}
\end{figure*}

These values have been used to build the differential reddening map in the direction of Terzan 6, over a FoV of about $200\arcsec \times 200\arcsec$, roughly corresponding to the FoV covered by the HST observations, with an angular resolution of only a few $0.1\arcsec$. The derived differential reddening map is shown in  Fig.~\ref{reddening_map}. As can be appreciated from the figure the reddening appears patchy on spatial scales of just a few arcseconds, with a highly reddened blob (dark color in the figure) located in the south-east region of the cluster. Another highly reddened structure can be distinguished diagonally crossing the north-west corner of the FoV as a sort of thin strip. Less absorbed regions (lighter color in the figure) are located in the south-west and the east portion of the cluster. Overall, the map graphically indicates a large amount of differential reddening, with $\delta E(B-V)$ ranging between -0.4 and 0.4. This corresponds to variation of differential absorptions of 2.1, 1.36, 0.63, and 0.26 mag in the V, I, J, and K filters, respectively.
The obtained reddening map has been used to correct the effect of the differential reddening from the observed CMDs.

\begin{figure*}[hbt!]
   \centering
 \includegraphics[width=12.5cm]{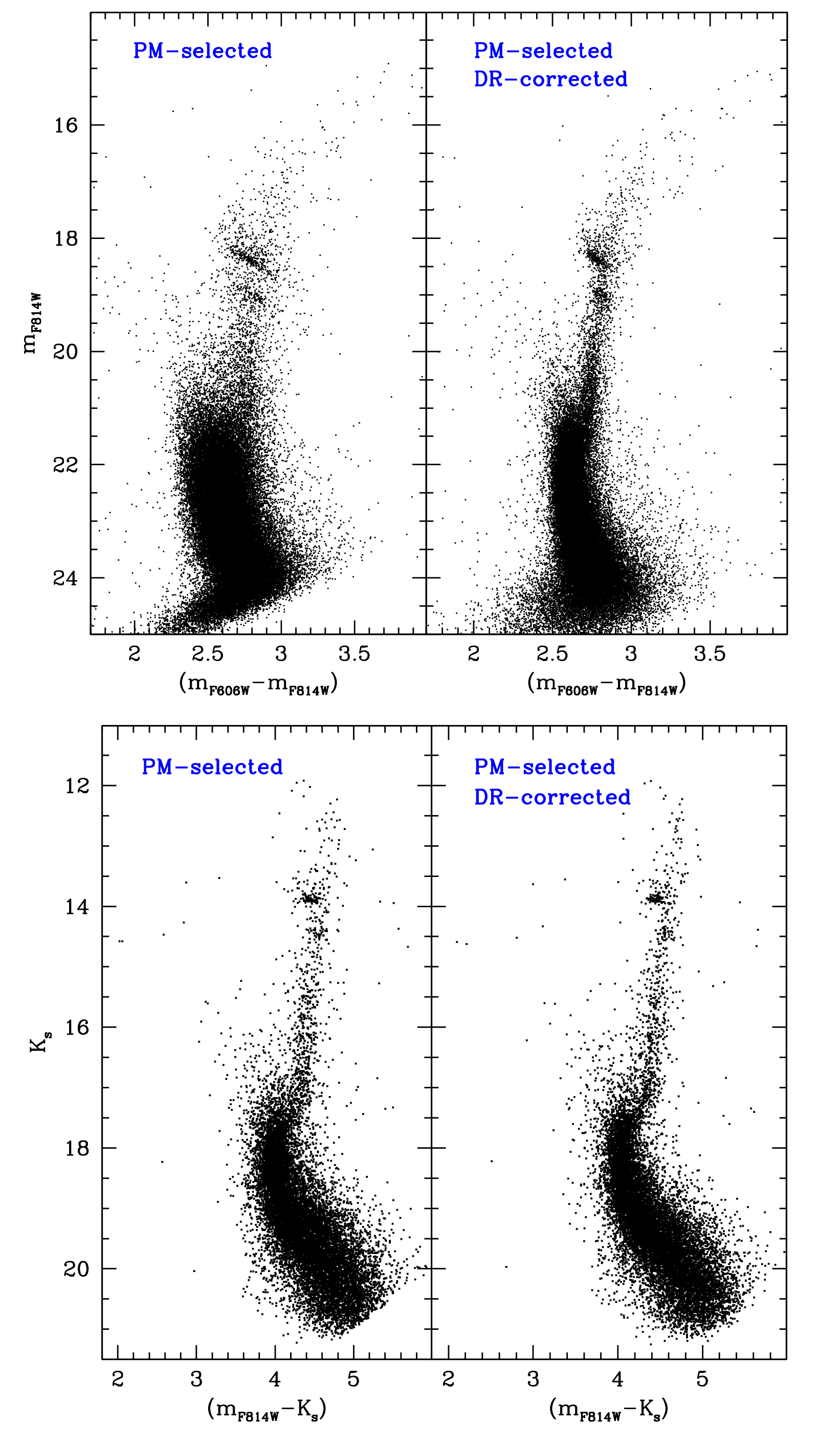}
   \caption{Optical (upper panels) and hybrid (bottom panels) CMD of the PM-selected member stars before (left panels) and after (right panels) the correction for differential reddening (DR in the plot).}
    \label{OPT_cmd_reddening}
\end{figure*}

The PM-selected and differential reddening-corrected optical CMD is shown in Fig.~\ref{OPT_cmd_reddening}. The effect of the correction for differential reddening is clearly visible, especially at the level of the red clump and the RGB bump, which both appear strongly elongated in the reddening direction non-corrected CMD. In the following analyses, the adopted magnitudes are always corrected for differential reddening. 

\section{Results}
\subsection{Center, density profile and structural parameters}
\label{Determination of the structural parameters}
The first step for the determination of the density profile is the estimate of the system's center of gravity ($C_{\rm grav}$). In this case it has been derived by using resolved star counts, thanks to the high spatial resolution of the data used for the analysis. With respect to integrated light studies, resolved star counts offer the main advantage of avoiding observational biases generated by the possible presence of a few bright stars that can offset the surface brightness peak from the true center position.
To determine $C_{\rm grav}$ we calculated the barycenter position of the stars selected within a given magnitude range and within a given radial distance, starting from a first-guess center \citep{harris+96}. Then, the procedure is iteratively repeated by adopting as new center the barycenter position calculated in the previous step, and convergence is reached when the difference between two successive estimates is smaller than $0.01\arcsec$ \citep[see, e.g.][]{montegriffo+95, miocchi+13}. The analysis was performed on stars selected with five radial distances (namely, 5, 10, 15, 20, 25 arcsec) and three faint magnitude limits ($m_{F814W} < 21, 21.5, 22$), thus providing 15 values of converge. The choice of the radial distances is set so that a gradient in the cluster density profile is sampled. A decrease of the density profile starts to be detectable around the core radius ($r_c$), hence the confidence radii have been chosen larger than the literature value of $r_c$ (namely, $3\arcsec$; \citealp{harris+96}). The faint magnitude limits are chosen to avoid spurious fluctuations due to photometric incompleteness, and we also adopted a bright cut (at $m_{F184W} = 16$) to exclude stars affected by saturation problems. The adopted magnitude limits also guarantee that the stars selected for the center's determination have approximately the same mass. The final result is obtained by averaging the 15 barycenter positions, and the associated error is the standard deviation of the mean. The obtained coordinates of $C_{\rm grav}$ are $\alpha_{\rm J200}$ = $17^{h}$$ 50^{m} $$ 46.9^{s}$, $\delta_{\rm J200}$ =$-31^{\circ}$ $16\arcmin$ $30.56\arcsec$, with an error of $0.44\arcsec$ on $\alpha$ and $0.18\arcsec$ on $\delta$. The new determination of the cluster center is very different from the one reported in the literature: it is located at $\approx 7\arcsec$ from the one quoted in \citet{harris+96}, toward the east and just $\sim 0.8\arcsec$ to the north.
   
\begin{figure}
    \centering
    \includegraphics[width=8cm]{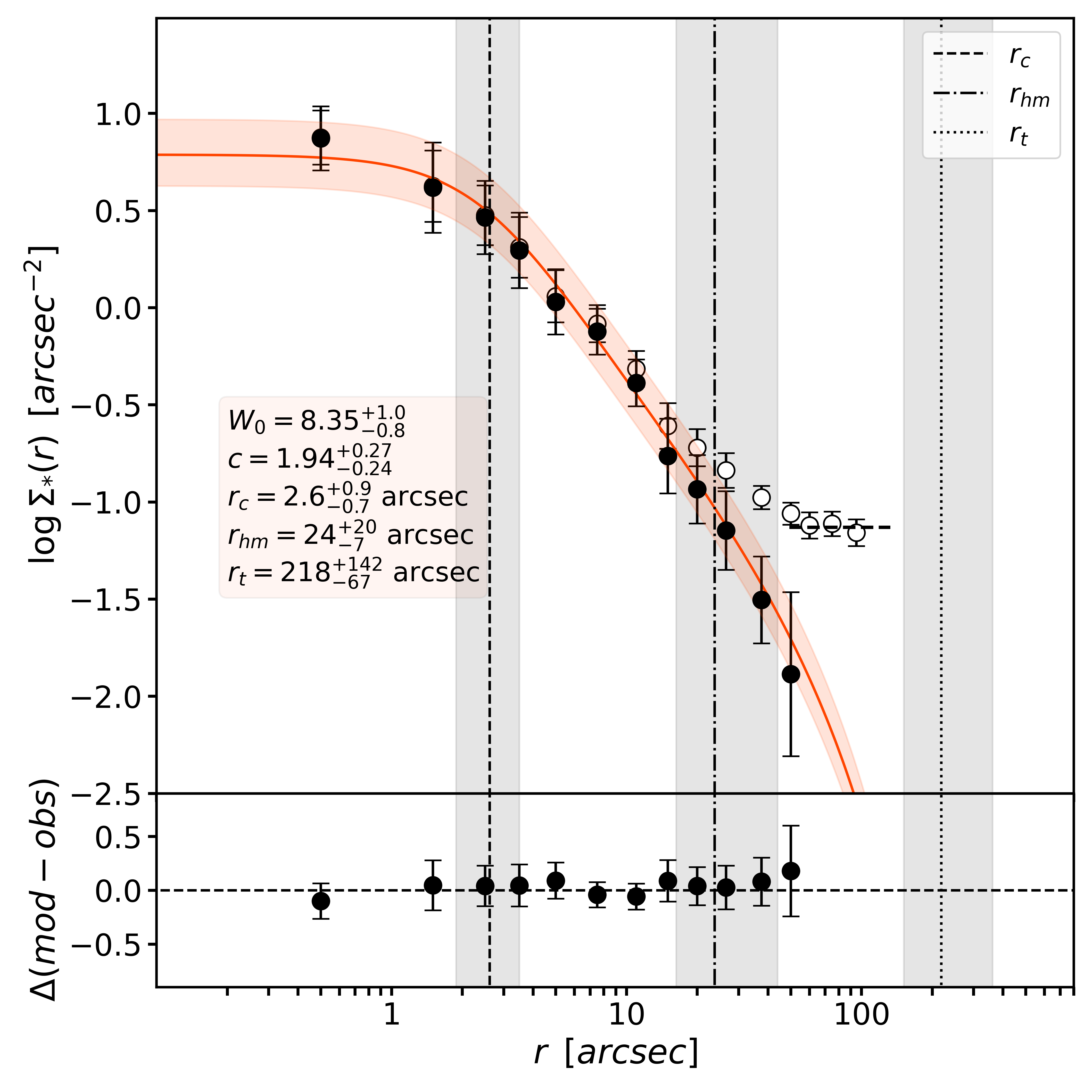}
    \caption{Projected density profile of Terzan 6 derived from star counts in concentric annuli around $C_{\rm grav}$ (empty circles). The horizontal dashed line indicates the Galactic field density, which has been subtracted from the observed points (empty circles) to obtain the background-subtracted profile, shown as filled circles. The red solid line represents the best-fit King model to the cluster's density profile, with the red stripe indicating the $\pm 1 \sigma$ range of solutions. The vertical lines denote the positions of the core radius (dashed line), half-mass radius (dot-dashed line), and tidal radius (dotted line), with their respective $1 \sigma$ uncertainties highlighted by gray stripes. The values of the main structural parameters obtained from the fitting process are also labeled (see details in the text).} 
    \label{dens_prof}
\end{figure}

Once $C_{\rm grav}$ has been obtained, we determined the radial density profile from resolved star counts \citep[see, e.g.,][for a detailed description of the procedure]{miocchi+13}. We split the FoV into concentric rings of different sizes chosen to ensure good statistics, and we determined the star density (number of stars divided by the ring area) within each of them. To avoid incompleteness problems, only stars with $m_{F814W}<20.8$ have been used. The projected density profile resulting from this operation is shown in Fig.~\ref{dens_prof} through empty circles. In the outermost regions, at about $r> 50\arcsec$, we note a flattening, with density values that remain constant at about $-1.13$ stars ${\rm arcsec}^{-2}$, due to the contribution of Galactic field stars. We determined the average level of the background as the mean of the densities of the 3 outermost rings. This value was subtracted from the observed profile, thus providing the true radial density distribution of the cluster (see the black circles in Fig.~\ref{dens_prof}). As expected, the field decontamination significantly affects the external portion of the profile, leaving the central one essentially unchanged.

To estimate the structural parameters of Terzan 6, we followed the procedure explained in \citet[see also \citealp{giusti+24,deras+24,deras+23,pallanca+23}]{raso+20}. It consists of a Monte Carlo Markov Chain (MCMC) fitting of the observed profile with \citet{king+66} models, assuming flat priors for the fitting parameters (namely the central density, the King concentration parameter $c$, and $r_c$) and a $\chi{^2}$ likelihood. The best-fit profile thus derived is shown in Fig.~\ref{dens_prof}. It is characterized by $r_c = 2.6_{-0.7}^{+0.9}$ arcsec, $c = 1.94_{-0.26}^{+0.24}$, which corresponds to a dimensionless central potential $W_0=8.35_{-0.8}^{+0.9}$, half-mass radius $r_{hm} = 24_{-7}^{+19}$ arcsec and tidal radius $r_t=219_{-67}^{+132}$ arcsec. The core radius is well consistent with that quoted in  \citet[][$r_c=3\arcsec$]{harris+96}. The concentration parameter is smaller than the literature one \citep[$c=2.5$ in][]{harris+96}, but still indicates an advanced stage of internal dynamical evolution for this cluster.

\subsection{Distance, reddening and age} 
\label{Distance and Age determination}
\begin{figure}
    \centering
    \includegraphics[width=8cm]{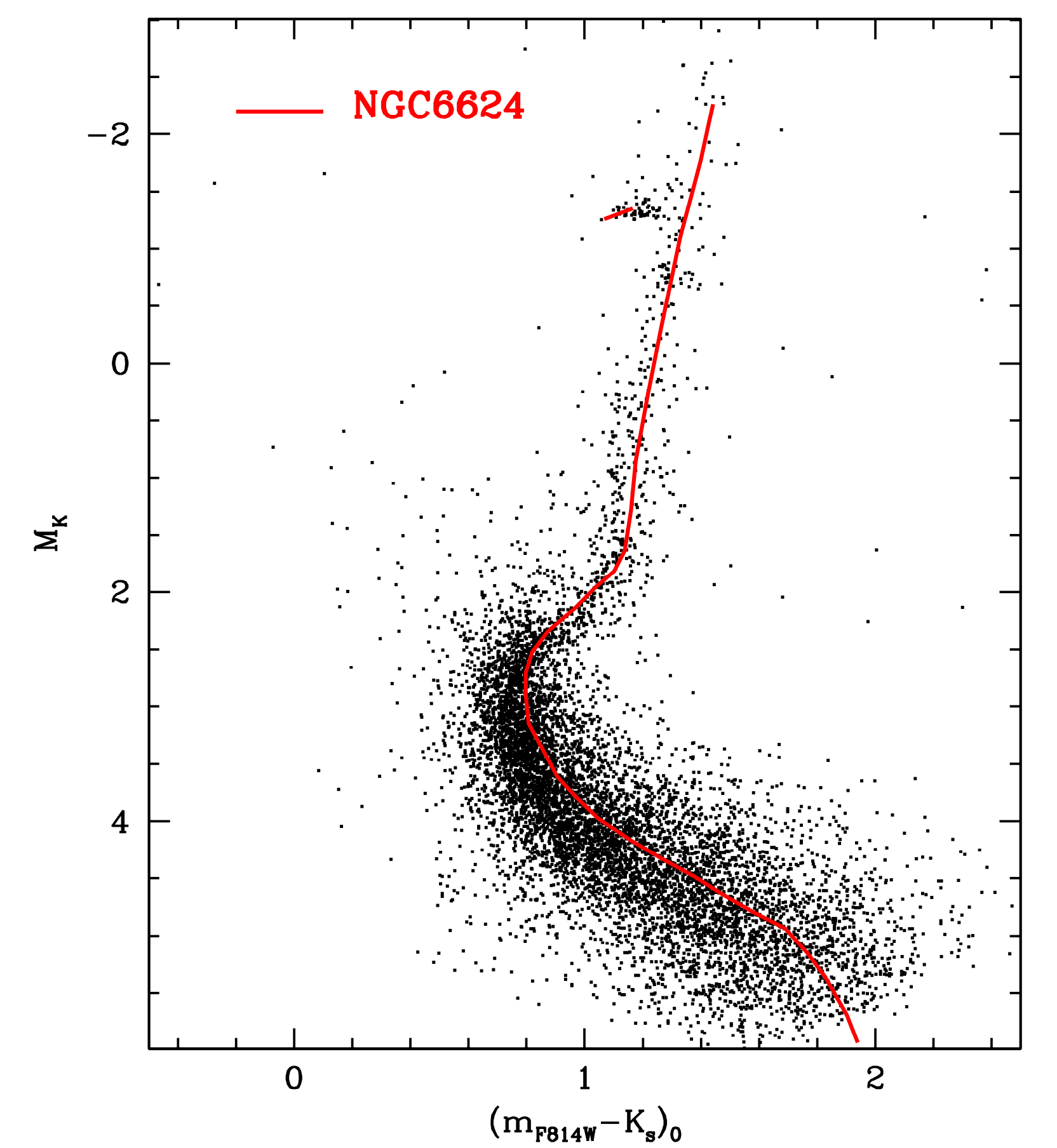}
    \caption{CMD of Terzan 6 in the absolute plane (black dots), with the MRL of NGC 6624 superposed as a red line.}
    \label{fig6624}
\end{figure}

\begin{figure*}
    \centering
    \includegraphics[width=15cm]{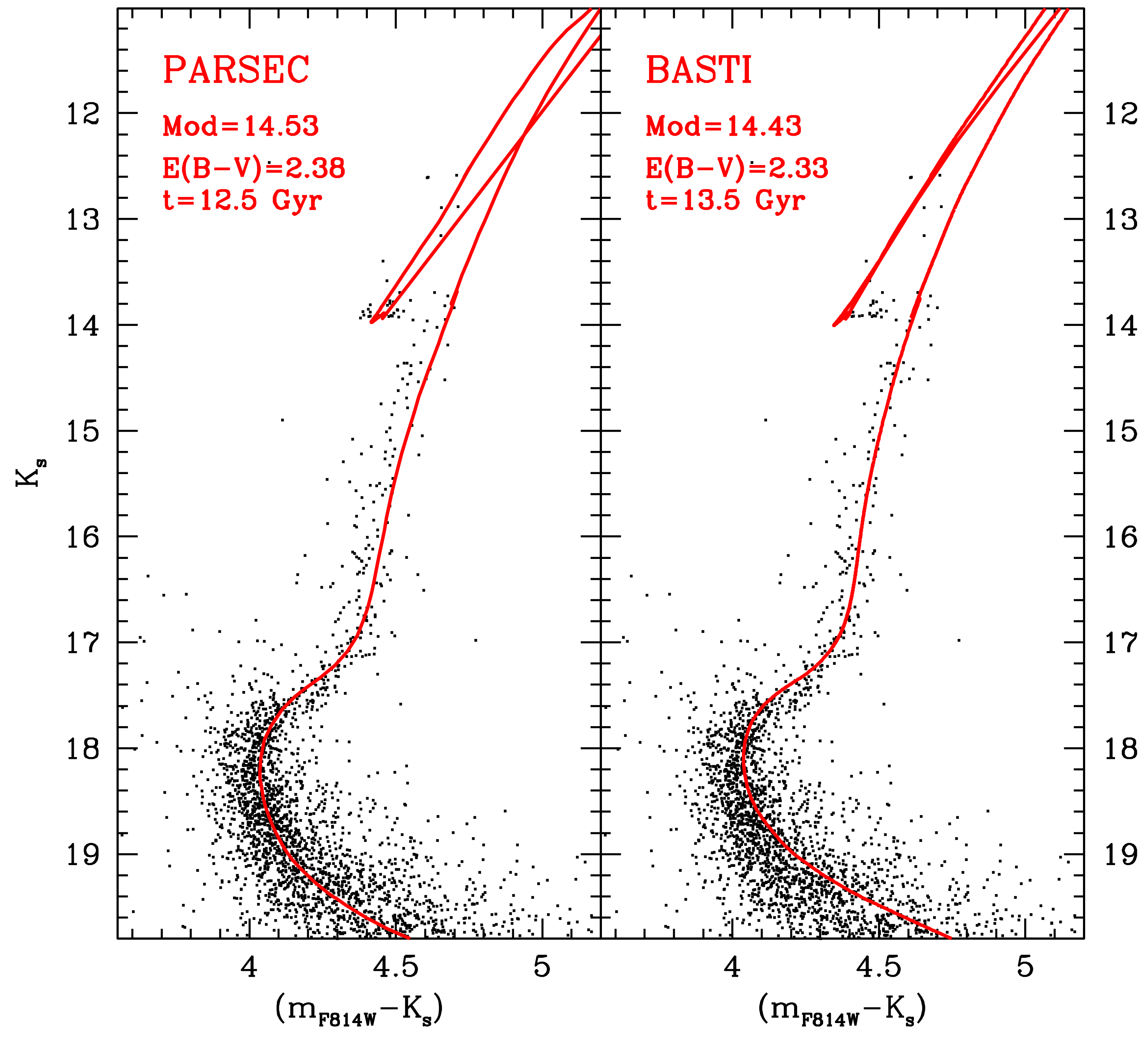}
    \caption{Differential reddening corrected and PM-selected ($K_s$, $m_{F814W}-K_s$) CMD of Terzan 6 with over-plotted the isochrones that best reproduce simultaneously the HB level and the MS-TO, for the PARSEC and the BASTI models (left and right panel, respectively). The values of the distance modulus, reddening and age required to optimize the match with the data are also labeled.}
    \label{combined_isochrones}
\end{figure*}

We obtained a first estimate of the distance modulus and mean color excess of Terzan 6 through the comparison with NGC 6624, a well-studied GC of similar metallicity ([Fe/H]$=-0.69$; \citealp{valenti+11}) with well-constrained foreground extinction $E(B-V)=0.28$ and true distance modulus $(m-M)_0=14.50\pm0.10$ \citep[see also \citealp{saracino+16} and \citealp{baumgardt+21}]{harris+96}.
To this aim, we shifted the PM-selected and differential reddening corrected CMD of Terzan 6 onto that of NGC 6624 reported into the absolute plane. We found that the optimal match is obtained by adopting  $E(B-V)= 2.37 \pm 0.05$ and a true distance modulus $(m-M)_0=14.45 \pm 0.15$ (see Figure \ref{fig6624})
Starting from these first-guess values, we then refined both the estimates and determined the age of Terzan 6 via isochrone fitting of the differential reddening corrected and PM-selected hybrid ($K_s$, $m_{F814W}-K_s$) CMD, which provides the best compromise between large wavelength baseline and good photometric accuracy at the level of the MS-TO. We also limited the analysis to the stars sampled in the radial range $5\arcsec<r<25\arcsec$, which optimize the definition of the MS-TO region.
The theoretical models used for this analysis are the \textit{BASTI} \citep{pietrinferni+04} and the \textit{PARSEC} \citep{bressan+12} isochrones\footnote{See, respectively, \href{http://basti-iac.oa-abruzzo.inaf.it/isocs.html}{basti-iac.oa-abruzzo.inaf.it/isocs.html} and \href{http://stev.oapd.inaf.it/cgi-bin/cmd}{stev.oapd.inaf.it/cgi-bin/cmd}}. We assumed [Fe/H]=$-0.65$, [$\alpha$/Fe]$=0.4$ and helium abundance Y=0.247, and used models also including overshooting and mass loss along the RGB \citep[see][\citeyear{origlia+07}]{origlia+02}.
We retrieved models calculated for ages ranging between 10 to 14 Gyr in both the ACS/WFC and the 2MASS photometric systems.
To find the best matching of the isochrones with the data we allowed the distance modulus and the reddening to vary within the 1-$\sigma$ error of the estimated first-guess values. The result of this iteration is plotted in Fig.~\ref{combined_isochrones}, showing the BASTI and the PARSEC isochrones that best reproduce simultaneously the Horizontal Branch (HB) level (left panels) and the MS-TO region (right panel). Slightly different values of the distance modulus, reddening and age are required to optimize the match between the two sets of isochrones and the data: $(m-M)_0=14.53 \pm 0.1$, $E(B-V)= 2.38 \pm 0.05$, and $t=12.5 \pm 1$ Gyr for the PARSEC isochrones, $(m-M)_0=14.43 \pm 0.1$, $E(B-V)= 2.33 \pm 0.05$, and $t=13.5 \pm 1$ Gyr for the BASTI models. Thus, we assumed the average values as final estimates of these parameters for Terzan 6, obtaining a true distance modulus  $(m-M)_0=14.46 \pm 0.10$ ($d=7.8 \pm 0.3$ kpc), a color excess $E(B-V)= 2.36 \pm 0.05$ and an age $t=13 \pm 1$ Gyr, with conservative estimates of the associated errors.
The derived distance modulus appears to be larger than the literature values: $(m-M)_0=14.11$ \citep{harris+96}, $(m-M)_0=14.16 \pm 0.14$ \citep{fahlman+95}, $(m-M)_0=14.25$ \citep{barbuy+97}, $(m-M)_0=14.13$ \citep{valenti+07}. However, it is important to notice that those works assumed $R_V=3.12$, while here we are using the new determination ($R_V=2.85$) discussed in Section \ref{Differential reddening}. In terms of distance, our result is compatible within the errors with the mean of the literature values, $d=7.3^{+0.4}_{-0.3}$ kpc \citep{baumgardt+21}. 

\begin{figure*}[hbt!]
    \centering
    \includegraphics[width=10cm]{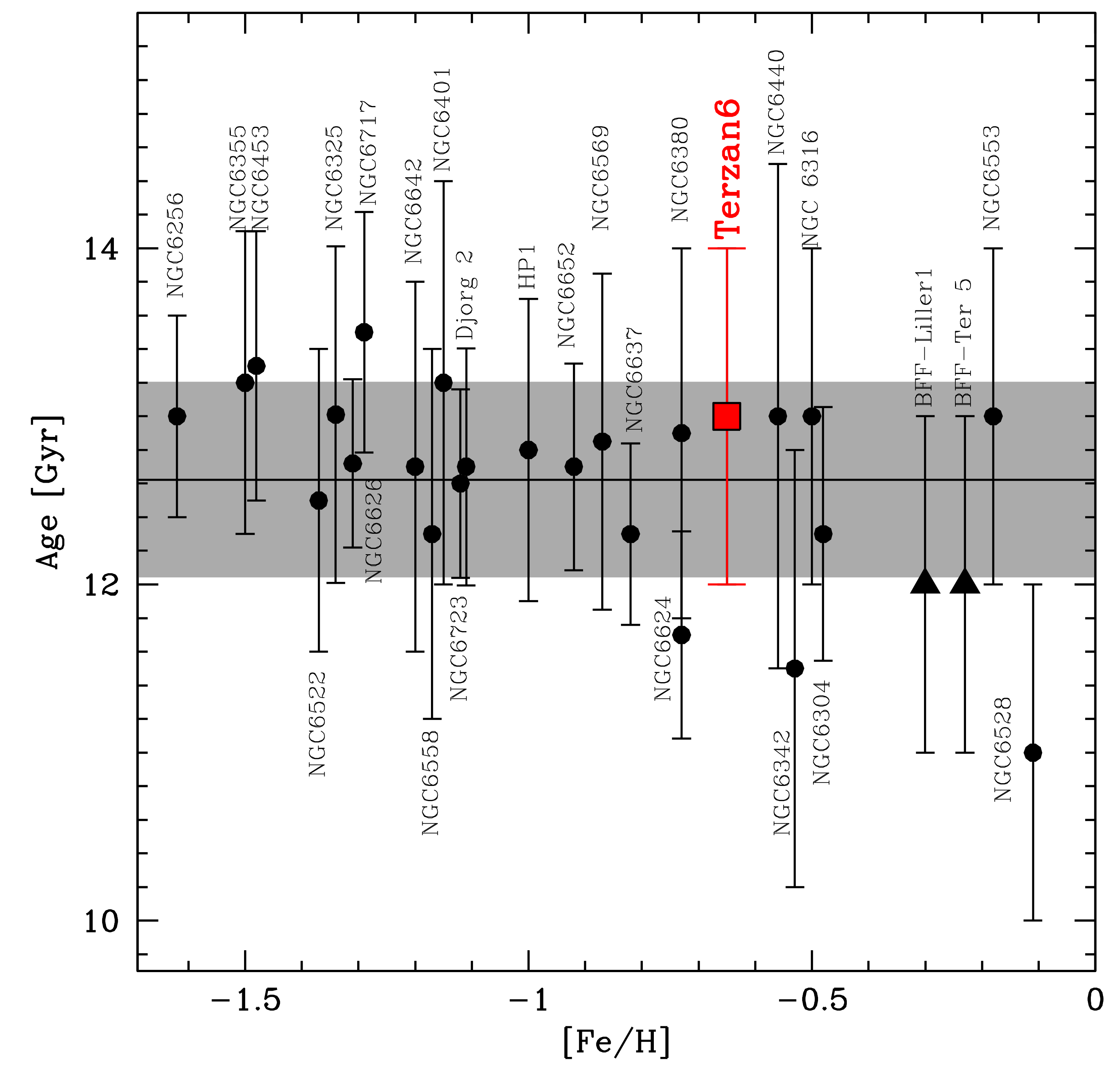}
	\caption{Age–metallicity distribution for the bulge GCs from the literature (black symbols) and for Terzan 6 (large red square). The literature data are mainly from \citet[][see their Figure 16]{saracino+19}, \citet[][see their Figure 12]{oliveira+20}, and \citet{cohen+21}, with the addition of a few recent age determinations for NGC 6440 \citep{pallanca+21}, NGC 6256 \citep{cadelano+20b}, and NGC6316 \citep{deras+23}. The age and metallicity of the oldest stellar population in the two suspected bulge fossil fragments (namely, Terzan 5 and Liller 1; \citealp{ferraro+09}, \citealp{ferraro+21}) are plotted as black triangles. The gray vertical strip marks the weighted average and the $1 \sigma$ uncertainty ($12.6\pm 0.6$ Gyr) of the entire sample.}
\label{age_met}
\end{figure*}

By exploiting the information about the age and the metallicity of the system, we can place Terzan 6 in the age-metallicity diagram (Fig.~\ref{age_met}), together with other Bulge GCs.
We can notice that the position of Terzan 6 is fully compatible within the errors with the average age of Bulge GCs, confirming an in-situ origin of this system (as already dynamically proved by \citealp{massari+19}).

\subsection{The RGB bump}
\label{sec:bump}
One of the most relevant evolutionary features characterizing the RGB is the so-called RGB bump. It flags the star luminosity at the moment when the hydrogen-burning shell crosses the hydrogen discontinuity left by the innermost penetration of the convective envelope (see \citealt{fusi+90, ferraro+91, ferraro+92, ferraro+99,ferraro+00}; see also the compilations by \citealt{zoccali+99, valenti+04}, and more recently by \citealt{nataf+13a}). Hence, the predicted luminosity of the RGB bump depends on all the parameters and physical processes affecting the penetration of convection deep into the stellar interior (for instance, the parameters affecting stellar opacity, such as the heavy element and helium abundances). In the case of a deep penetration, the hydrogen-burning shell is expected to cross the chemical discontinuity left by the convective envelope at an early stage of the RGB evolution, and the bump occurs at faint magnitudes along the RGB. Thus, a strong dependence is expected between the RGB bump luminosity and the overall metallicity of the stellar population.
From the observational point of view, the RGB bump can be easily identified via the detection of a local increase of star counts (a "bump in star counts") along the RGB differential luminosity function (e.g., \citealt{fusi+90, ferraro+99, ferraro+00}), due to a temporary hesitation in the star evolutionary path when crossing the hydrogen discontinuity.

\begin{figure*}[hbt!]
    \centering
    \includegraphics[width=10cm]{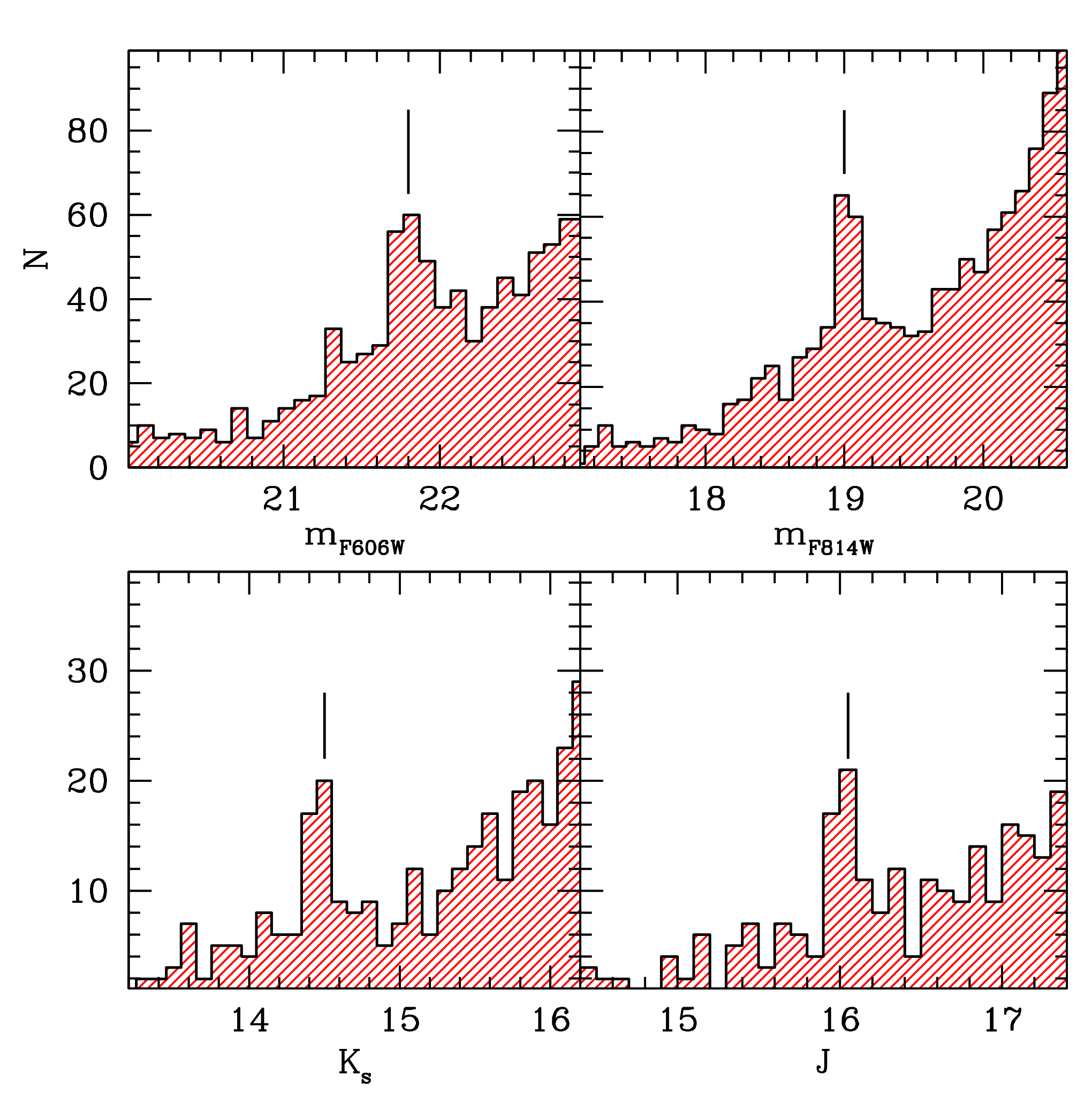}
	\caption{Differential luminosity function of RGB stars classified as cluster members. The detected peaks (marked by vertical black segments) are the RGB bump in the four photometric bands (see labels).}
\label{bump}
\end{figure*}

The high-quality CMDs obtained in this study offer the possibility to identify this feature in Terzan 6. Indeed, already from a visual inspection of Fig. \ref{OPT_cmd_reddening}, the location of the RGB bump is easily recognizable in the CMD as a small clump of stars along the RGB at slightly lower luminosity than the red clump. Fig. \ref{bump} shows the RGB differential luminosity function in different filters obtained from the PM-selected and differential reddening corrected CMDs shown in Fig. \ref{OPT_cmd_reddening}. The RGB bump is located at $m_{F606W}=21.80\pm0.05$, $m_{F814W}=19.00\pm0.05$, $J=16.05\pm0.05$, and $K_s=14.50\pm0.05$.

\begin{figure*}[hbt!]
    \centering
    \includegraphics[width=10cm]{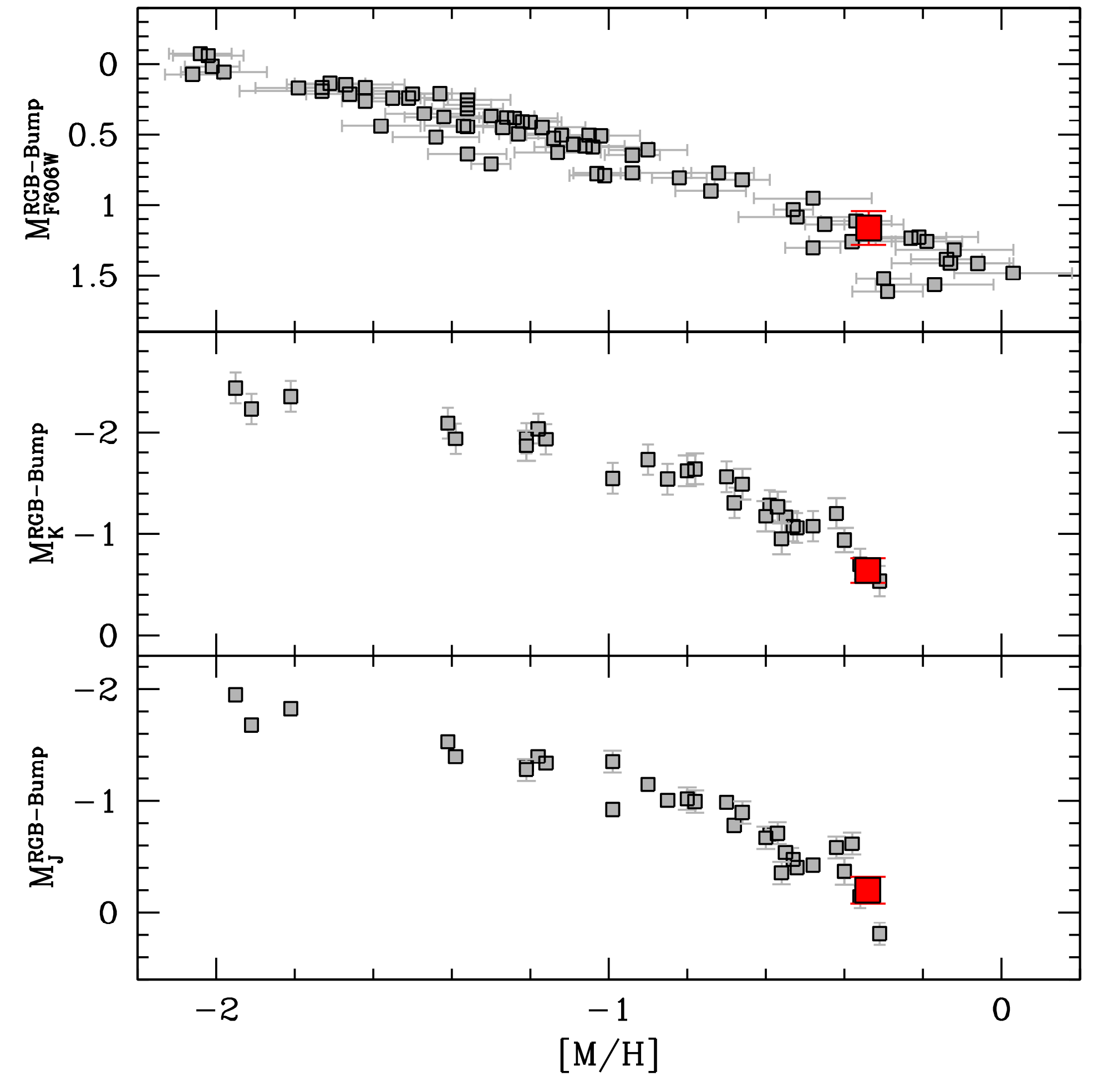}
	\caption{Absolute magnitude of the RGB bump in the $m_{F606W}$, $K$, and $J$ bands (from top to bottom) as a function of the cluster global metallicity [M/H]. The small gray squares are data from the literature: in the upper panel from \citealp{nataf+13a,pallanca+21,deras+24}, in the central and bottom panels from  \citealp{valenti+04,valenti+07,pallanca+21}. In all the panels the large red squares mark the location of the RGB bump estimated here for Terzan 6.}
\label{bump2}
\end{figure*}

Adopting the color excess and distance modulus derived in Section \ref{Distance and Age determination}, the absolute magnitude of the RGB bump is $M_{F606W}=1.16\pm0.12$,
$M_{F814W}=0.59\pm0.12$, $M_J=-0.20\pm0.12$, and
$M_{K_s}=-0.64\pm0.12$. As pointed out by \citealt{ferraro+99}, to evaluate the dependence of the RGB bump luminosity on the metallicity, one must consider the global metallicity ([M/H]), which includes the contribution from $\alpha-$elements and can be estimated from the relation \citep{salaris+93}: [M/H]=[Fe/H] $+\log(0.638 \times f_{\alpha}+0.362)$, where $f_{\alpha}=1 0^{[\alpha/Fe]}$. Thus, assuming [Fe/H]=$-0.65$ and [$\alpha$/Fe]$=0.4$, for Terzan 6 we obtain [M/H]$=-0.34$. In Fig. \ref{bump2} we show the comparison among the value obtained here for Terzan 6 to those derived for other Galactic GCs in the literature \citep{nataf+13a,pallanca+21,deras+24}, in the $V-$ and NIR bands. 
Indeed, the Terzan 6 determinations nicely fit into the relations defined by previous works.

\section{Summary and conclusions}
\label{Conclusions}
This work is set in the framework of using GCs as excellent
tracers of the early stages of the formation and evolution of the MW. In particular, it is part of a larger project aimed at obtaining a detailed photometric \citep{saracino+15, saracino+16, saracino+19,deras+23,deras+24,pallanca+19, pallanca+21,cadelano+20a,cadelano+23,ferraro+23}, chemical  \citep{crociati+23,fanelli+24a,fanelli+24b,deimer+24}, kinematic \citep{leanza+23,pallanca+23,libralato+22,ferraro+18} and chronological characterization of the population of Galactic Bulge clusters, which have been poorly studied so far because of the large foreground extinction and stellar density. To overcome these limitations, we used a combination of high-resolution, multi-epoch NIR and optical data sets acquired with GeMS/GSAOI at GEMINI, and with the ACS/WFC onboard the HST. Thanks to the high angular resolution offered by both data sets, and to the NIR sensitivity of the GeMS@GEMINI, it has been possible to perform the first detailed study of the stellar population hosted in the heavily reddened GC Terzan 6.
The derived CMDs span $\approx$ 10 magnitudes, allowing us to observe all the main evolutionary sequences from the red clump level, down to $\approx$ 3 mag below the MS-TO.
Thanks to the multi-epoch data set and the large temporal baseline, it was possible to derive relative PMs for 135134 stars in this highly contaminated stellar system. To do this, we exploited all 3 HST epochs and the GEMINI epoch. In particular, we computed the relative PMs by using 2 different epochs for 71201 stars, 3 different epochs for 53118 stars, and all 4 different epochs for 10815 stars. This analysis enabled a clear separation between cluster members and stars belonging to the Galactic field, thus making possible a detailed characterization of the stellar population of the system.
Then, we obtained a high-resolution differential reddening map in the direction of the cluster, finding that $\delta E(B-V)$ varies by $\approx 0.8$ mag in the FoV of our data sets. The reddening law in the direction of the system is non-canonical, with $R_V=2.85$.  
By using the sky positions of resolved stars, we obtained a new estimate of the center of the system, finding that it is very different (i.e., offset by $\sim 7\arcsec$ to the east with respect to the one quoted in the literature and obtained through surface brightness studies \citep{ter71, harris+96}.
We also determined the very first projected density profile obtained from resolved star counts. The analysis presented here indicates that Terzan 6 has experienced or is on the verge of experiencing the collapse of the core. 
By taking advantage of the PM-selected and differential reddening corrected CMDs, we estimated that Terzan 6 is located at a distance $d=7.8 \pm 0.3$ kpc from the Sun, and has an absolute age $t=13\pm 1 $ Gyr.

In summary, this work allowed the characterization of the stellar population properties of the Bulge GC Terzan 6 with a level of detail never reached before. It also shows the wealth of information that it is possible to obtain even in high-density and large extinction regions such as the Galactic Bulge, from the combination of NIR adaptive optics supported systems and optical space-based instruments.

\begin{acknowledgements}
This work is part of the project Cosmic-Lab at the Physics and Astronomy Department “A. Righi” of the Bologna University (\href{http://www.cosmic-lab.eu/Cosmic-Lab/Home.html}{www.cosmic-lab.eu/Cosmic-Lab/Home.html}). M.L. acknowledges funding from the European Union NextGenerationEU.
D.G. gratefully acknowledges the support provided by Fondecyt regular n. 1220264.
D.G. also acknowledges financial support from the Direcci\'on de Investigaci\'on y Desarrollo de
la Universidad de La Serena through the Programa de Incentivo a la Investigaci\'on de
Acad\'emicos (PIA-DIDULS).
S.V. gratefully acknowledges the support provided by Fondecyt Regular n. 1220264 and by the ANID BASAL project ACE210002.

\end{acknowledgements}

\end{document}